\documentclass[12pt,onecolumn,prb,showpacs]{revtex4}
\usepackage{graphicx,color}
\usepackage{amsmath}
\usepackage{longtable}
\begin{document}

\title{Ionization enhanced ion collection by a small floating grain in plasmas}

\author{Sergey A. Khrapak$^{1,2,}\footnote{Electronic mail: skhrapak@mpe.mpg.de}$ and Gregor E. Morfill}
\date{\today}
\affiliation{ $^1$Max-Planck-Institut f\"ur extraterrestrische Physik, D-85741 Garching,
Germany \\$^2$Joint Institute for High Temperatures, 125412 Moscow, Russia}

\begin{abstract}
It is demonstrated that the ionization events in the vicinity of a small floating grain can increase the ion flux to its surface. In this respect the effect of electron impact ionization is fully analogous to that of the ion-neutral resonant charge exchange collisions. Both processes create slow ion which cannot overcome grain' electrical attraction and eventually fall onto its surface. The relative importance of ionization and ion-neutral collisions is roughly given by the ratio of the corresponding frequencies. We have evaluated this ratio for neon and argon plasmas to demonstrate that ionization enhanced ion collection can indeed be an important factor affecting grain charging in realistic experimental conditions.
\end{abstract}

\maketitle

The problem of interactions between an object and surrounding plasma is a fundamental issue with many applications including technological and astrophysical topics, methods of plasma diagnostics,  fusion related aspects, etc. It is especially important in the context of complex (dusty) plasmas -- plasmas containing charged micron-size particles (grains).~\cite{Book,FortovPR}
The key element of basic plasma-particle interactions is the process of particle charging, since the charge is responsible for essentially every phenomenon observable in complex plasmas.~\cite{KhrapakCPP} Not surprisingly, this process has been extensively investigated employing various experimental, theoretical and simulation methods. In the last decade many works have been focused on the effect of ion-neutral collisions in the vicinity of the grain.~\cite{Zobnin,Lampe,FortovPRE,RatynskaiaPRL,KhrapakPRE,Vaulina,Hutchinson,Diachkov,Zobnin1,KM2008,Gatti}
It has been demonstrated that in the weakly collisional regime, typical for complex plasma experiments, ion-neutral collisions can greatly enhance ion collection by the grain, which leads to a significant reduction in the grain charge (absolute magnitude) compared to the collisionless situations.

The purpose of the present paper is to discuss another mechanism affecting the grain charge in complex plasmas, which has apparently been overlooked in previous studies. This is associated with the ionization events in the vicinity of the grain, which have an effect very similar to that of ion-neutral collisions. Both processes can be described using essentially equivalent physical arguments, which allows for an easy comparison between them. We show, that ionization-related effect can be dominant for sufficiently high (but realistic) electron temperatures.


We start with a brief discussion of the collisional enhancement of the ion flux to a small floating grain in the weakly collisional regime. Following arguments of Lampe~\cite{Lampe,Note1} it is instructive to think of a sphere of a radius $R_0$ such that inside this sphere the (attractive) interaction between an ion and the grain is sufficiently strong. The distance $R_0$ can be roughly defined from the condition that the energy of ion-grain interaction for $r\leq R_0$ is higher than the average ion kinetic energy after a collision (exact definition is not important for the present purposes). If an ion undergoes a collision (especially of resonant charge exchange type) with a neutral within $r<R_0$, then a fast ion is removed and a slow ion is created. This new low-energy ion has very low probability to overcome the attraction of the grain and escape back into the plasma bulk. It will eventually reach the grain surface, either directly (low angular momentum) or through subsequent collisions (high angular momentum). Thus, essentially every (charge exchange) collision of an ion within $r<R_0$ results in ion collection by the grain. The collisional contribution to the collected ion flux can be estimated as the influx of ions through the spherical surface of radius $R_0$ ($\simeq\sqrt{8\pi} R_0^2 n_i v_{T_i}$) multiplied by the probability to experience a collision within this sphere ($\sim R_0/\ell_i\ll 1$ in the weakly collisional regime). Here $n_i$ is the ion density far from the grain, $v_{T_i}=\sqrt{T_i/m_i}$ is the ion thermal energy, and $\ell_i$ is the ion mean free path with respect to collisions with neutrals. We get therefore $\Delta J_{\rm coll}\simeq \sqrt{8\pi} n_i v_{T_i} (R_0^3/\ell_i)$. If we add this to the collisionless orbital motion limited (OML) ion flux,~\cite{OML,Allen} we obtain an approximate expression derived by Lampe~\cite{Lampe}
\begin{equation}
J_i\simeq \sqrt{8\pi}a^2 n_i v_{T_i}\left[1+z\tau+(R_0^3/a^2\ell_i)\right],
\end{equation}
where $z=|Q|e/aT_e$ is the reduced grain charge and $\tau=T_e/T_i$ is the electron-to-ion temperature ratio.

Note that introducing the effective ion-neutral collision frequency, $\nu_{in}=v_{T_i}/\ell_i$, the collisional contribution to the ion flux can be rewritten as $\Delta J_{\rm coll}\simeq \sqrt{8\pi}R_0^3n_i \nu_{in}$. This is by about $20\%$ larger than a simplified, but physically transparent estimate
\begin{equation}\label{Jcoll}
\Delta J_{\rm coll}\simeq (4\pi/3)R_0^3n_i \nu_{in},
\end{equation}
stating that the collisional correction to the ion flux is roughly the number of ion-neutral collisions per unit time inside the sphere of radius $R_0$. In general, $n_i(r)$ can experience significant perturbations in the vicinity of the grain, but from the derivation it is obvious that the unperturbed value $n_i$ should be used in Eq. (\ref{Jcoll}).


The arguments applied above to evaluate the effect of ion-neutral charge exchange collisions are equally relevant to the electron impact ionization events. When neutral atom is ionized within $r<R_0$ from the grain, a new slow ion is created, which has almost no chances to escape from the potential well. It will, therefore, very likely reach the grain surface. From the point of view of contribution to the collected ion flux, resonant charge exchange collisions and ionization events are essentially equivalent. In full analogy with Eq.~(\ref{Jcoll}) we can write
\begin{equation}\label{JI}
\Delta J_{\rm I}\simeq (4\pi/3)R_0^3n_e\nu_{\rm I},
\end{equation}
where $\nu_I$ is the ionization frequency. The meaning of $n_e$ in Eq.~(\ref{JI}) is the mean electron density inside the sphere of radius $R_0$ surrounding the grain. However, since perturbations in $n_e(r)$ are essentially localized to a close proximity of the grain (region where $e|\phi(r)|\gtrsim T_e$ corresponds to $r\lesssim az$ for the Coulomb distribution of electrical potential $\phi(r)$ near the grain) this value practically coincides with the unperturbed electron density far from the grain. Using the quasineutrality condition $n_i\simeq n_e$ in the unperturbed plasma we immediately see that the relative magnitude of collisional and ionization effects is simply given by the ratio of the correspomding frequencies:
\begin{equation}
\Delta J_{\rm I}/\Delta J_{\rm coll}\equiv {\mathcal K}\simeq \nu_{\rm I}/\nu_{\rm in}.
\end{equation}

Let us estimate the expected magnitude of ${\mathcal K}$ under typical laboratory condition. For conventional situations with $T_e < I$, the behavior of the ionization cross section near the ionization threshold is of main interest. Here, to a good accuracy, the cross section exhibits linear growth with energy $\sigma_{\rm I}\simeq \sigma_0(\varepsilon/I-1)$, where $I$ is the ionization potential and $\varepsilon$ is the electron energy. The ionization frequency is obtained by averaging this cross section over the electron energy distribution function, $\nu_{\rm I}= n_n\langle\sigma_{\rm I}v_e\rangle$, where $n_n$ is the neutral gas density. For Maxwellian electrons we get~\cite{Raizer}
\begin{equation}
\nu_{\rm I}= \sqrt{8/\pi} n_n\sigma_0 v_{T_e} \left(1+2T_e/I\right)\exp(-I/T_e),
\end{equation}
where $v_{T_e}=\sqrt{T_e/m_e}$ is the electron thermal velocity.

\begin{figure}
\centering
\includegraphics[width= 10 cm]{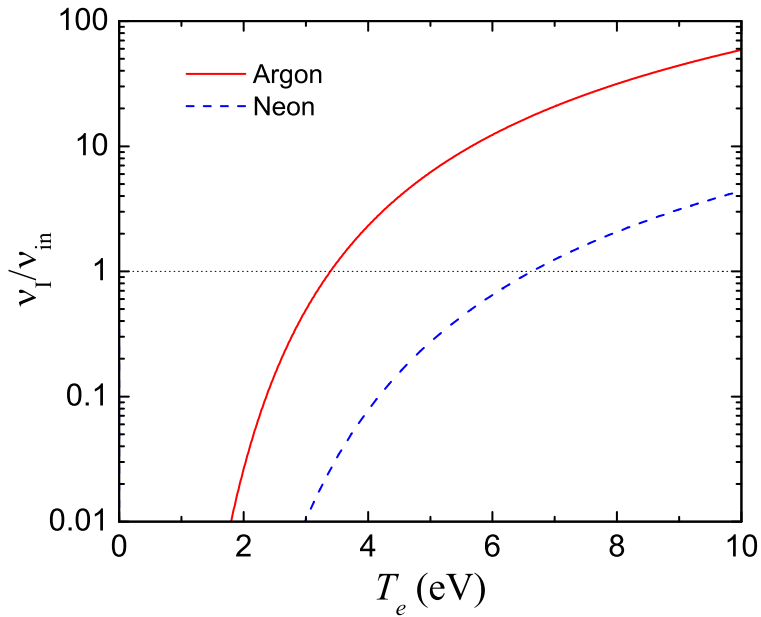}
\caption{The ratio of ionization and ion-neutral collision frequencies ${\mathcal K}=\nu_{\rm I}/\nu_{\rm in}$ in a plasma of argon (solid curve) and neon (dashed curve) as a function of the electron temperature $T_e$. The parameters used in calculation are summarized in Table \ref{table}, ions are at room temperature. The horizontal dotted line corresponds to ${\mathcal K}=1$.}\label{Fig1}
\end{figure}

The effective ion-neutral collision frequency is $\nu_{\rm in}\simeq v_{T_i}n_n\sigma_{\rm in}$, where for slow ions in their parent gases $\sigma_{\rm in}$ is mainly determined by the resonance charge exchange. The ratio ${\mathcal K}$ can thus be written as
\begin{equation}\label{ratio}
{\mathcal K}\simeq \sqrt{\frac{8}{\pi}}\left(\frac{\sigma_0}{\sigma_{\rm in}}\right)\sqrt{\frac{T_e}{T_i}\frac{m_i}{m_e}}\left(1+\frac{2T_e}{I}\right)\exp\left(-\frac{I}{T_e}\right).
\end{equation}
For a given gas, the quantities $\sigma_0$, $\sigma_{\rm in}$, $m_i/m_e$ and $I$ are fixed. Equation (\ref{ratio}) implies that ${\mathcal K}$ depends only on the ion and electron temperatures. If, in addition, we assume that ions are at room temperature $T_i\simeq 0.03$ eV (as usual for most of gas discharges involved in complex plasma investigations), the ratio ${\mathcal K}$ becomes a function of the electron temperature only. We have evaluated this function for neon and argon plasmas using the parameters summarized in Table \ref{table}. The results are shown in Fig.~\ref{Fig1}.
It turns out that ${\mathcal K}$ is a very sharply increasing function of $T_e$. For argon it becomes comparable to unity for $T_e\gtrsim 3$ eV, while for neon this happens for $T_e\gtrsim 6.5$ eV. Both values are not unrealistically high for conventional gas discharges used in complex plasma research. For example, electron temperatures in the range between $\simeq 8$ eV and $\simeq 6$ eV have been reported for a dc discharge in neon (PK-4 Project) operating at a current of 1 mA in the pressure range between 20 and 150 Pa.~\cite{Usachev} Even higher electron temperatures (up to the first excitation potential of 16.6 eV in neon) have been discussed in connection to the dusty plasma structures levitating in the head of a standing striation in a dc glow discharge,~\cite{Lipaev} but in this case the electron energy distribution function can be far from Maxwellian. The electron temperatures in the range $T_e\simeq 3-4$ eV have been measured in the International Microgravity Plasma Facility (IMPF) chamber (an rf discharge) filled with argon gas at pressures between 15 and 100 Pa.~\cite{Klindworth} Simulations of the PK-3 Plus complex plasma chamber (an rf discharge), now operational onboard the International Space Station, reveal electron temperatures around $4$ eV in argon and $5-6$ eV in neon, for typical discharge conditions without the grains.~\cite{Hubertus} The temperature can further increase when the dust grains are injected into the discharge.~\cite{Land} Thus, the ionization-related effects can be expected to affect grain charging under many realistic conditions.

Finally, let us discuss an important related issue. Present estimate assumes Maxwellian electron energy distribution function (EEDF). This is certainly not always the case in practical situations. For example, distributions close to the Druyvesteyn shape have been documented in a recent numerical study of electron drift in a dc electric field in neon.~\cite{Mayorov} Although the electron flux collected by the grain is virtually independent of the EEDF (provided the average electron energy remains the same), the ionization rate is affected considerably.~\cite{KhrapakPoP2010} Depletion of the high-energy tail lowers the ionization rate and suppress the effect of ionization on grain charging. Thus, proper attention should be given to the exact shape of EEDF in each concrete situation.

To conclude, we have shown that ionization events in the vicinity of a small floating grain in a plasma can enhance ion flux to its surface. The increase in the ion flux results in a decrease of the absolute magnitude of the floating potential (and, therefore, charge) of the grain. The effect becomes significant when the electron temperature is sufficiently (but not unrealistically) high and is sensitive to the electron energy distribution function.

\newpage

\begin{table}
\caption{\label{table} Values of parameters used to estimate the ratio of ionization and ion-neutral collision frequencies. The electron impact ionization cross section scales $\sigma_0$ are estimated from the experimental results of Ref.~\onlinecite{Fletcher}. The ion-neutral cross section scales $\sigma_{\rm in}$ are taken from Ref.~\onlinecite{Jovanovich}.}
\begin{ruledtabular}
\begin{tabular}{lllll}
Gas      & $\sigma_0$ (cm$^2$) & $\sigma_{\rm in}$ (cm$^2$)  & $m_i/m_e$ & $I$ (eV)   \\ \hline
Neon     & $3.5\times 10^{-17}$ & $1\times 10^{-14}$ & $3.7\times 10^4$ & $21.6$   \\
Argon    & $3.2\times 10^{-16}$ & $2\times 10^{-14}$ & $7.5\times 10^4$ & $15.8$   \\
\end{tabular}
\end{ruledtabular}
\end{table}

\end{document}